\documentclass[12pt,groupedaddress,superscriptaddress,onecolumn,a4,preprint]{article}

\usepackage{lipsum}

\usepackage[left=1.5cm,top=2cm, right=1.5cm, bottom=1.5cm]{geometry}
\usepackage{amsmath,amssymb}
 \usepackage{graphicx,float}
 \usepackage{subfigure}
 \usepackage{dcolumn}
\usepackage{bm}
\usepackage{mathrsfs}
\usepackage{txfonts}
\usepackage{CJK}
\usepackage[amssymb]{SIunits}
\usepackage{epsfig}
\usepackage{lipsum}
\usepackage{color}
\usepackage{textcomp}
\usepackage[table]{xcolor}
\usepackage{multirow}
\usepackage[T1]{fontenc}
\usepackage{times}
\usepackage{authblk}
\usepackage{placeins}
\usepackage[numbers,sort]{natbib}

\begin{document}

%
%
%
%
\title{Deterministic generation of bright multicolor entanglement from optomechanical systems}

\author[1]{\hskip 1cm Keyu Xia \thanks{keyu.xia@mq.edu.au} }

\affil[1]{ARC Centre for Engineered Quantum Systems, 

Department of Physics and Astronomy, Macquarie University, NSW 2109, Australia}

\date{\today}

\allowdisplaybreaks[2]


\maketitle

\begin{abstract}
  Entangled continuous variable (CV) Gaussian states with different wavelengths plays a central role in recent CV-based approaches to quantum network, quantum information processing and quantum metrology. Typically, experiments demonstrating CV entanglement exploit the optical parametric frequency down conversion. Due to the probabilistic nature of photon pair generation, the entanglement involving the post-selection of photonic qubits is limited to at most three colors. Here We theoretically present a scheme for the deterministic generation of entanglement among bright multicolor CV Gaussian states from an optomechanical system  using existing experimental technologies. In our scheme an optical frequency comb is input into an optomechanical resonator and then the amplified optomechanical coupling makes multipartite entanglement among them. Our scheme overcomes the limitation of usable frequency of entangled CVs in frequency conversion process. It can be extended to generate multipartite entanglement between orthogonal modes with a single frequency or between microwave and optical CVs, or even among a microwave frequency comb. This optomechanical device can be integrated on a chip.
\end{abstract}

\newpage

Multipartite entanglement of continuous variable (CV) cluster states is not only of fundamental scientific interest \cite{ThreeColorEnt1}, but also the key ingredient for quantum information technologies such as universal quantum computation \cite{NielsenClusterState1,NielsenClusterState2,RevModPhys.84.621-GaussianReview}, quantum metrology \cite{QuantMetrology1,QuantMetrology2}, gravitational wave detection \cite{Gravition}, and even quantum network of clocks \cite{QuantClock}.
 
A number of different techniques for the generation of entanglement in the CV regime have been proposed and experimentally realized. Entangled CV cluster states are created by combining squeezed Gaussian states generated from optical parametric oscillator (OPO) on beamsplitter but is limited to one frequency \cite{MultipartiteEnt2,PhysRevLett.111.250403-SqueezingComb, NatPhys11.167-SqueezingComb,NatCommun3.1026-SqueezingComb}. Signal and idler photons generated from OPOs during the frequency down conversion (FDC) are naturally entangled but limited to two color \cite{MultipartiteEnt1,TwoColorEnt1,TwoColorEnt2,TwoColorEnt3}. This FDC-based technique is only extended up to entangle three-color CVs  \cite{ThreeColorEnt1,ThreeColorEnt2} due to two limitations: the probabilistic nature of FDC and the available frequency reduced by half in each FDC. We note that Four-wave mixing technique has been demonstrated to be able to create quantum correlation among three-color CVs but entanglement is not clear\cite{FWMJietai}.

 To date, entangled CV cluster states has only been generated up to two colors in frequency domain via the probabilistic FDC recently \cite{MultipartiteEnt1, TwoColorEnt2,TwoColorEnt3}, or in spatial modes \cite{NatCommun3.1026-SqueezingComb}, or in time domain via optical group delay \cite{MultipartiteEnt2}. However, the wavelength or color of entangled CV states is limited by available optical nonlinear crystals or materials.
%
In this work, we propose a method using a multimode optomechanical system to create entanglement among multicolor CV cluster state. We apply an optical frequency comb (OFC) of coherent laser field to drive the cavity modes on-resonance. In our configuration, the reflected CV Gaussian states off cavity with at least ten colors around each cavity mode frequency are entangled. Our scheme overcomes the fundamental limitation of mode number of entangled CVs in frequency domain.\newline

\begin{figure}
 \centering
 \includegraphics[width=0.6\linewidth]{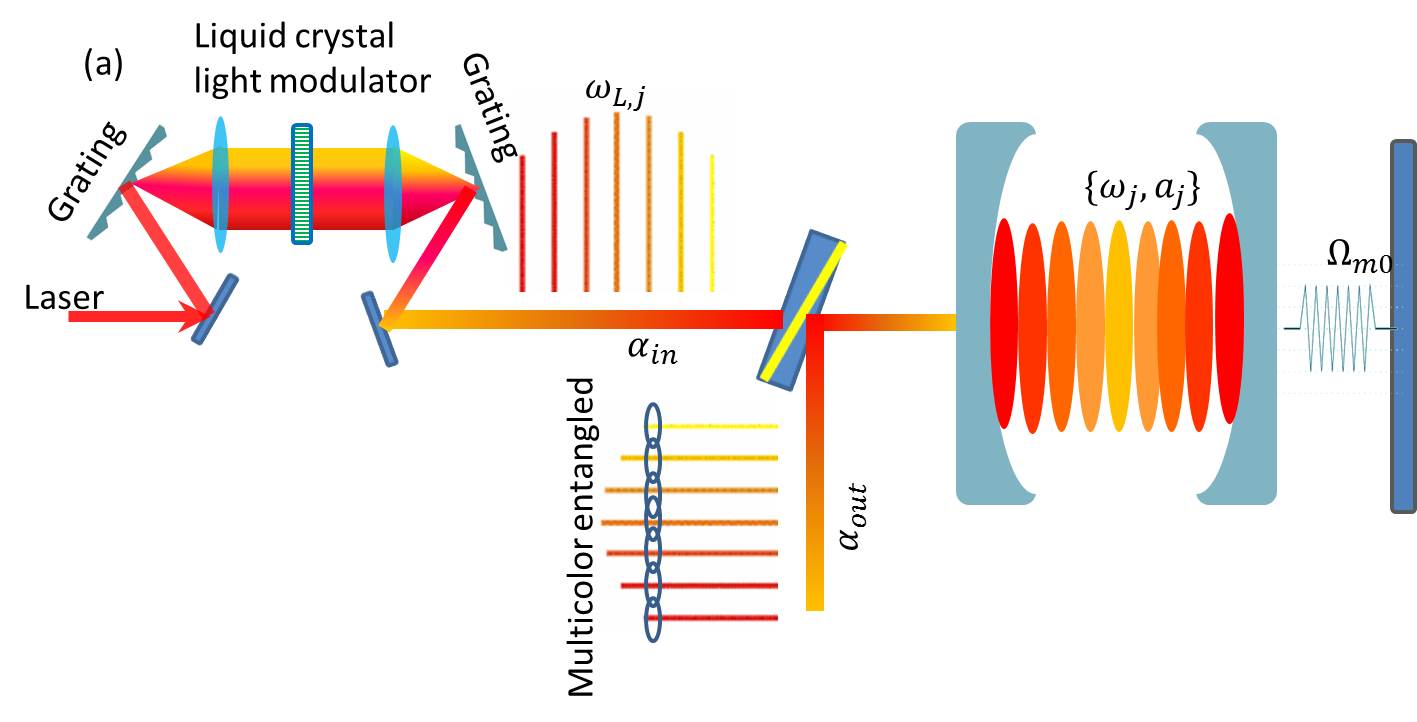}
 \includegraphics[width=0.25\linewidth]{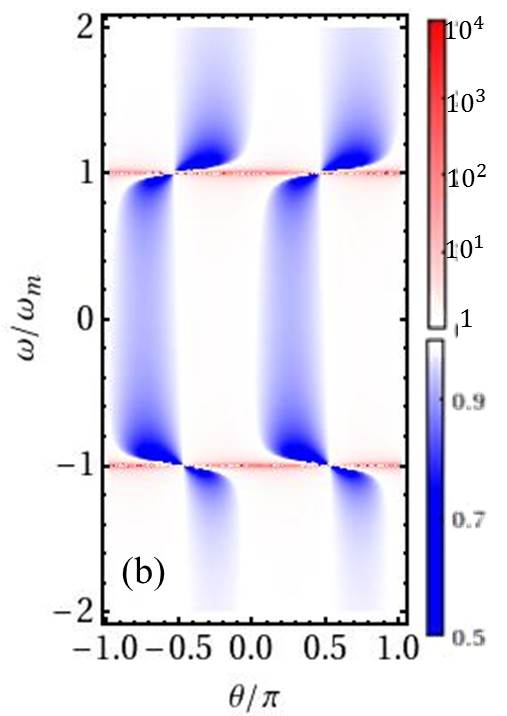}\\
 \caption{(a) Schematic setup for generation of multicolor entanglement in optomechanical systems. The spectrum decomposited by the fist optical grating from an optical frequency comb is modulated by a liquid crystal light modulator (LCLM) and then is combined by the second optical grating. This modulated frequency comb is input into a multimode optomechanical resonator and subsequently is reflected off the optomechanical system. The optomechanical resonator creates entanglement between the spectral lines (with frequency $\omega_{\text{L},j}$) of the input optical frequency comb. (b) Entanglement of two modes when $G=0.3\Omega_{\text{m}0}$. The plot shows $V^{(2)}_{12}/2$. Other parameters are $\Delta=0, \Omega_{\text{m}0}=0.1\kappa, Q_m=10^6,\bar{n}_\text{th}=10^3,\kappa_\text{i}=0$.}\label{fig:system}
\end{figure}

\noindent{\bf Results}\\
\noindent{\bf Theoretical description.} 
The setup for generation of multicolor CV entanglement is illustrated schematically in Fig. \ref{fig:system}(a). The optical frequency comb is filtered and then incidents to the first optical grating. The spectra decomposited by this grating is modulated by a liquid crystal light modulator (LCLM) with fast response to the amplitude and phase of each spectral line \cite{MultipartiteEnt1}, and then is combined by the second optical grating. The modulated optical frequency comb incidents into the optomechanical resonator supporting optical cavity multimode, which has resonance frequency $\omega_j$ and intrinsic loss rate of $\kappa_{\text{i},j}$ for the $j$\emph{th} mode. The $j$\emph{th} CV mode of optical frequency comb at frequency $\omega_{\text{L},j}$ drives the corresponding $j$\emph{th} cavity mode with an amplitude of $\varepsilon_j$, and an external coupling rate of $\kappa_{\text{e},j}$. One of two mirrors of optomechanical resonator oscillates with frequency $\Omega_{\text{m}0}$ and its motion decays with a rate of $\gamma_{\text{m}0}$. The mechanical motion of this movable mirror couples to the $j$\emph{th} cavity mode with a rate of $g_{\text{om},j}$. The optical fields reflected off the OMR is isolated from the input fields by a highly reflective beam splitter. Using this setup, we can create entanglement among the multicolor CV modes of OFC.

Under the rotating-wave approximation of the driving, the Hamiltonian describing the interaction between the optical modes and the mechanical motion is \cite{PhysRevLett.99.093901}:
\begin{equation}
 \begin{split}
  H & =\Omega_{\text{m}0} b^\dag b + \sum_j \Delta^\prime_j a_j^\dag a_j + \sum_j g_{\text{om},j}a_j^\dag a_j (b^\dag + b) \\
 & + i\sum_j \sqrt{2\kappa_{\text{e},j}} (\varepsilon_j a^\dag_j - \varepsilon_j^* a_j) \;,
 \end{split}
\end{equation}
with the detuning $\Delta^\prime_j = \omega_j -\omega_{\text{L},j}$. Here $a_j$ and $b$ are the annihilation operators of the $j$\emph{th} cavity mode and the mechanical motion, respectively. $g_{\text{om},j}$ is the zero-point optomechanical coupling rate of the $j$\emph{th} cavity mode. Driven by multi CV modes in OFC, the cavity modes and the mechanical motion can reach the steady state \cite{PhysRevLett.99.093901,PhysRevLett.99.093902}, 
 $\bar{a}_j  = \langle a_j\rangle_{ss} = \frac{\sqrt{2\kappa_{\text{e},j}}\varepsilon_j}{i\Delta_j + \kappa_j} \;,
 \beta  = \langle b\rangle_{ss} = - \frac{i\sum_j g_{\text{om},j} |\bar{a}_j|^2}{i\Omega_{m0}+ \gamma_{m0}} \approx - \frac{1}{\Omega_{m0}}\sum_j g_{\text{om},j} |\bar{a}_j|^2$,
where $\Delta_j = \Delta^\prime_j + 2 g_{\text{om},j} \beta$ and $\kappa_j = \kappa_{\text{e},j} + \kappa_{\text{i},j}$. 
In this work we are interested in the noise included in the output fields. We define the $X$ and $Y$ quadratures of the output fields for various angle $\theta$ as $X_{\text{out},j}(\omega)= \frac{e^{-i\theta} a_{\text{out},j}(\omega) + e^{i\theta} a^\dag_{\text{out},j}(\omega)}{\sqrt{2}}$ and $Y_{\text{out},j}(\omega)  = \frac{e^{-i\theta} a_{\text{out},j}(\omega) - e^{i\theta} a^\dag_{\text{out},j}(\omega)}{i\sqrt{2}}$.  By substituting $a_j \rightarrow \bar{a}_j + a_j$, $b \rightarrow \beta + b$ and the output $a_{\text{out},j} \rightarrow \bar{a}_{\text{out},j} + a_{\text{out},j}$, we linearize the operators to a weak fluctuation displaced by a strong coherent shift $\bar{a}_j$, $\beta$ and $\bar{a}_{\text{out},j}$ \cite{PhysRevLett.99.093901,PhysRevLett.99.093902}, respectively. Normally, $|\bar{a}_{\text{out},j}|^2$, i.e. the photon number of output field, is much larger than unity, implying a bright output beam. Correspondingly, the fluctuation in the $X$ and $Y$ quadratures are $\Delta X_{\text{out},j}(\omega)= \frac{e^{-i\theta} a_{\text{out},j}(\omega) + e^{i\theta} a^\dag_{\text{out},j}(\omega)}{\sqrt{2}}$ and $\Delta Y_{\text{out},j}(\omega)  = \frac{e^{-i\theta} a_{\text{out},j}(\omega) - e^{i\theta} a^\dag_{\text{out},j}(\omega)}{i\sqrt{2}}$ \cite{NatureOneModeSqueezingPainter}. After linearization, the Hamiltonian describing the dissipation and noise from environment becomes \cite{QuntNoise1,QuntNoise2}:
\begin{equation}
 H^\prime = \Omega_{\text{m}0} b^\dag b + \sum_j \Delta_j a^\dag_j a_j + \sum_j (G_j a^\dag_j + G_j^* a_j)(b^\dag + b) \;,
\end{equation}
with $G_j=g_{\text{om},j} \bar{a}_j$. Obviously, one can tune $G_j$ to be real by controlling the phase of driving, $\varepsilon_j$. Throughout the investigation below, we assume all the amplified coupling rates $G_j$ real for simplicity. We now study the quantum fluctuation in the output using the quantum Langevin equations (QLEs)  \cite{QuntNoise1,QuntNoise2}:
\begin{subequations}\label{eq:QLE}
\begin{align}
 \dot{a}_j & = -(i\Delta_j + \kappa_j)a_j - i G_j(b^\dag + b)  + \sqrt{2\kappa_{\text{e},j}}a_{\text{in,e},j}(t)  + \sqrt{2\kappa_{\text{i},j}}a_{\text{in,i},j}(t) \;,\\
  \dot{b} & = -(i\Omega_{m0}+\gamma_{m0})b - i\sum_j G_j (a_j^\dag +a_j) + \sqrt{2\gamma_{\text{m}0}} b_\text{in}(t) \;,
\end{align}
\end{subequations}
where the optical vacuum noise entering the cavity through the external optical coupling and the intrinsic optical loss channels are included by the random fluctuating inputs $a_{\text{in,e},j}(t)$  and $a_{\text{in,i},j}(t)$, respectively. $b_\text{in}(t)$ describes the mechanical noise applying to the mechanical resonator. $\gamma_{m0}$ is the decay rate of mechanical motion and therefore gives mechanical quality factor $Q_m=\Omega_{m0}/\gamma_{m0}$.
To solve the QLEs in frequency domain, we define the Fourier and inverse-Fourier transformations $\hat{A}(\omega)  =\frac{1}{\sqrt{2\pi}} \int_{-\infty}^\infty d\omega e^{i\omega t} \hat{A}(t)$ and 
 $\hat{A}(t)  = \frac{1}{\sqrt{2\pi}} \int_{-\infty}^\infty d\omega e^{-i\omega t} \hat{A}(\omega)$ \cite{NatureOneModeSqueezingPainter,OptoRevAAClerk,OptoRevKippenberg}. Thus we have the correlation for noises $\langle a_{\text{in,e},j}(\omega)a^\dag_{\text{in,e},l}(\omega^\prime) \rangle=\delta_{jl} \delta(\omega+\omega^\prime)$, $\langle a_{\text{in,i},j}(\omega)a^\dag_{\text{in,i},l}(\omega^\prime) \rangle=\delta_{jl} \delta(\omega+\omega^\prime)$, where $\delta_{jl}$ ($\delta(\omega+\omega^\prime)$) is the discrete Kronecker (Dirac) delta function. Here we assume that the occupancy of cavity modes due to the thermal environment is negligible, $\langle a^\dag_{\text{in,e(i)},j}(\omega)a_{\text{in,e(i)},j}(\omega^\prime) \rangle=0$. But the optical vacuum noises from external coupling is uncorrelated to that from the intrinsic loss, i.e $\langle a_{\text{in,e},j}(\omega)a^\dag_{\text{in,i},l}(\omega^\prime) \rangle=0$ for any $j$ and $l$. We also have $\langle b_\text{in}(\omega)b^\dag_\text{in}(\omega^\prime)= (\bar{n}_\text{th}+1)\delta(\omega+\omega^\prime)$ and $\langle b^\dag_\text{in}(\omega)b_\text{in}(\omega^\prime)= \bar{n}_\text{th}\delta(\omega+\omega^\prime)$, where $\bar{n}_\text{th}= (e^{\hbar \Omega_{\text{m}0}/K_BT}-1)^{-1}$ is the thermal occupancy of mechanical resonator at temperature $T$. But $\langle b_\text{in}(\omega)b_\text{in}(\omega^\prime)= 0$ and  $\langle b^\dag_\text{in}(\omega)b^\dag_\text{in}(\omega^\prime)= 0$. The solution to the QLEs, Eq. \ref{eq:QLE}, in the Fourier domain is given by \cite{NatureOneModeSqueezingPainter,OptoRevAAClerk,OptoRevKippenberg}
\begin{subequations} \label{eq:ab}
 \begin{align}
  a_j(\omega) & = \sqrt{2\kappa_{\text{e},j}} \chi_{R,j}(\omega) a_{\text{in,e},j}(\omega) + \sqrt{2\kappa_{\text{i},j}} \chi_{R,j}(\omega) a_{\text{in,i},j}(\omega) -iG_j \chi_{R,j}(\omega) \left[b^\dag(\omega)  + b(\omega)\right] \;, \\
  b(\omega) & = \frac{\sqrt{2\gamma_{\text{m}0}} b_\text{in}(\omega) -i\sum_j G_j\left[ a^\dag_j(\omega) + a_j(\omega) \right]}{i(\Omega_{\text{m}0}-\omega) + \gamma_{\text{m}0}} \;,
 \end{align}
\end{subequations}
where the optical cavity susceptibility for the $j$\emph{th} mode is $\chi_{\text{R},j}(\omega) = 1/[i(\Delta_j - \omega) + \kappa_j]$. The mechanical susceptibility $\chi_\text{m}(\omega)$ connecting the mechanical response to the environmental noise and the optical vacuum fluctuations incident on the optical cavity takes the form $\chi^{-1}_m(\omega) = i (\Omega_{m0} -\omega) + \gamma_{m0} + \sum_j G_j^{2}\left[  \chi_{R,j}(\omega) - \chi^*_{R,j}(-\omega) \right]$. 

The fluctuations in the output fields can be calculated from the input-output relation \cite{PhysRevA.30.1386,PhysRevA.31.3761} $a_{\text{out},j}(\omega) = - a_{\text{in,e},j}(\omega) + \sqrt{2\kappa_\text{e}} a_j(\omega)$. Then we have the output noise correlations \cite{NatureOneModeSqueezingPainter,PRXOneModeSqueezing,ThreeColorEnt1}
\begin{subequations}\label{eq:FullCorrel}
 \begin{align}
 \langle \Delta X_{\text{out},j}(\omega) \Delta X_{\text{out},l}(-\omega)\rangle  =& \frac{1}{2}  \delta_{jl} + 2 \sqrt{\kappa_{e,j}\kappa_{e,l}} G_l G_j |F_m(\omega)|^2 \zeta_j(\omega)\zeta^*_l(\omega)\sum_n \kappa_{n} G_n^2 |\chi_{R,n}(\omega)|^2 \\ \nonumber
 &  + 2\gamma_{m0}\sqrt{\kappa_{ej}\kappa_{el}} G_j G_l \zeta_j(\omega)\zeta^*_l(\omega)\left[|\chi_m(\omega)|^2 (\bar{n}_m +1) + |\chi_m(-\omega)|^2 \bar{n}_m\right]  \\ \nonumber
& - 2\sqrt{\kappa_{ej}\kappa_{el}} G_lG_j \Re\left[ \left(2\kappa_j\chi_{R,j}(\omega)-1 \right) F_m^*(\omega) \zeta^*_l(\omega)\chi^*_{R,j}(\omega) e^{-i\theta}\right]  \;, \\
\langle\Delta  Y_{\text{out},j}(\omega) \Delta Y_{\text{out},l}(-\omega)\rangle  = & 
 \frac{1}{2}  \delta_{jl}  + 2 \sqrt{\kappa_{e,j}\kappa_{e,l}} G_l G_j |F_m(\omega)|^2 \varXi_j(\omega)\varXi^*_l(\omega)\sum_n \kappa_{n}  G_n^2 |\chi_{R,n}(\omega)|^2   \\ \nonumber
 & +2\gamma_{m0}\sqrt{\kappa_{ej}\kappa_{el}} G_j G_l \varXi_j(\omega)\varXi^*_l(\omega)\left[|\chi_m(\omega)|^2 (\bar{n}_m +1) + |\chi_m(-\omega)|^2 \bar{n}_m\right]   \\ \nonumber
& + 2 \sqrt{\kappa_{ej}\kappa_{el}} G_lG_j \Re\left[ \left(2\kappa_j\chi_{R,j}(\omega)-1 \right) F_m^*(\omega) \varXi^*_l(\omega)\chi^*_{R,j}(\omega) e^{-i\theta}\right] \;,
\end{align}
\end{subequations}
where $F_\text{m}=\chi^*_\text{m}(-\omega)-\chi_\text{m}(\omega)$,  $\zeta_j(\omega) =  e^{i\theta}\chi^*_{Rj}(-\omega) -e^{-i\theta}\chi_{Rj}(\omega)$ and $\varXi_j(\omega) =  e^{i\theta}\chi^*_{Rj}(-\omega) +e^{-i\theta}\chi_{Rj}(\omega)$. $\Re[x]$ means the real part of number $x$. 

One of the criteria for analyzing multipartite entanglement of CV modes is Duan criteria \cite{DuanCriterion1} and its extension \cite{DuanCriterion2} written directly in terms of these correlations, as sums of variances $V_{jl}$ between modes $j$ and $l$:
\begin{equation} \label{eq:FullV}
\begin{split}
 V^{(M)}_{jl}(\omega) = & \langle \left(\Delta X_j(\omega) - \Delta X_l(\omega)\right)\left(\Delta X_j(-\omega) - \Delta X_l(-\omega)\right)\rangle  \\
 & + \langle \left(\Delta Y_j(\omega) + \Delta Y_l(\omega)\right)\left(\Delta Y_j(-\omega) + \Delta Y_l(-\omega)\right)\rangle \geq 2 \;,
\end{split}
\end{equation}
for $j\neq l$. It suffices to demonstrate $M$-partite entanglement as long as the above inequality is violated for $j<l$ and $j,l \in \{1,2,\cdots, M\}$ \cite{ThreeColorEnt1,EntComb}. The degree of violation indicates the degree of entanglement. Note that the inequality Eq. \ref{eq:FullV} is not necessary optimal for searching the largest entanglement but its violation is a sufficient criterion for inseparability.

Next we present a simply closed formula for estimation of the correlation and variance for the case all cavity modes are identical, i.e. $\Delta_j=\Delta, G_j=G, \kappa_{\text{e},j}= \kappa_\text{e}$, and $\kappa_{\text{i},j}= \kappa_\text{i}$. Although bipartite squeezed states have been proposed in optomechanical systems by applying the red-detuned and blue-detuned multitone driving simultaneously \cite{SqueezingMatthew}, it is hard to be scaled for multipartition entanglement. Here we take the scheme using a single zero or small detuned driving recently demonstrated in experiments for squeezing a single optical mode in optomechanical systems \cite{PRXOneModeSqueezing,NatureOneModeSqueezingPainter,NatureOneModeSqueezingDaniel}. Therefore, we take $\Delta\approx 0$. We define $\delta_\pm = \Omega_{\text{m}0}\pm \omega$ and $\delta^{-1}=\delta_-^{-1} + \delta_+^{-1}$, and assume that $|\delta|\gg \gamma_{\text{m}0}$. Under this condition, for cavity modes with identical decay and optomechanical coupling rates, we have 
\begin{subequations}\label{eq:SmpCorrel}
 \begin{align}
   \langle \Delta X_{\text{out},j}(\omega)\Delta X_{\text{out},l}(-\omega)\rangle = & \frac{1}{2}\delta_{jl} + \eta M\left(\frac{\tilde{\Gamma}_\text{meas}}{2\delta}\right)^2(1-\cos2\theta) \\ \nonumber
 &  + \eta \frac{\tilde{\Gamma}_\text{meas}}{\delta} \left[\Omega_m\delta\frac{\bar{n}_m}{Q_m} (\delta_-^{-2}+\delta_+^{-2})(1-\cos2\theta) + \frac{\sin2\theta}{2} \right] \;,\\
  \langle \Delta Y_{\text{out},j}(\omega)\Delta Y_{\text{out},l}(-\omega)\rangle = & \frac{1}{2}\delta_{jl} + \eta M\left(\frac{\tilde{\Gamma}_\text{meas}}{2\delta}\right)^2(1+\cos2\theta) \\ \nonumber
 & + \eta \frac{\tilde{\Gamma}_\text{meas}}{\delta} \left[\Omega_m\delta\frac{\bar{n}_m}{Q_m} (\delta_-^{-2}+\delta_+^{-2})(1+\cos2\theta) - \frac{\sin2\theta}{2} \right]  \;,
 \end{align}
\end{subequations}
with $\eta = \kappa_e/\kappa$ and $\tilde{\Gamma}_\text{meas}=\frac{4G^2}{\kappa} \frac{\kappa^2}{\omega^2+\kappa^2}$. According to Eq. \ref{eq:SmpCorrel}, if one quadrature, e.g. $X_{\text{out}}$, is correlated, another quadrature, $Y_{\text{out}}$, will be anticorrelated. As shown in Fig. \ref{fig:system}(b), the minimal variance is obtained around $|\theta|=\pi/2, 3\pi/2$. However the frequency width of entanglement is small (also see supplementary material). In contrast, the optomechanical system has a broad width where entanglement is realized when $|\theta\pm \pi/2|=pi/4$. Next we focus on $|\theta|=\pi/4$. In the region of $|\omega|\sim \Omega_m$ such that $\delta^2 (\delta_-^{-2}+\delta_+^{-2})\approx 1$, the variance as sums of correlations is
\begin{equation} \label{eq:SmpV}
  V^{(M)}_{jl}(\omega) \approx 2 + \eta M\left(\frac{\tilde{\Gamma}_\text{meas}}{\delta}\right)^2  + \eta \frac{\tilde{\Gamma}_\text{meas}}{\delta}(4\frac{\Omega_m}{\delta}\frac{\bar{n}_m}{Q_m} -2) \;.
\end{equation}
The first term is the Duan's bound. The entanglement between the $j$\emph{th} and $l$\emph{th} modes requires $|\delta| > M\tilde{\Gamma}_\text{meas}/2 + \frac{\Omega_m\bar{n}_m}{Q_m}$, but the minimal variance, $V_\text{min} \approx 2- \frac{\eta\tilde{\Gamma}_\text{meas}}{M \tilde{\Gamma}_\text{meas} + 4 \Omega_{\text{m}0} \frac{\bar{n}_m}{Q_m}} > 2- \eta/M$, is available at the optimal frequency $|\delta_\text{opt}|= M\tilde{\Gamma}_\text{meas} + 4 \Omega_\text{m0}\frac{\bar{n}_\text{th}}{Q_\text{m}}$. Interestingly, the largest achievable entanglement is independent of the optomechanical coupling strength, but is limited by the number of involved cavity modes.

According to our analysis above, the perfect coupling regime \cite{NatureOneModeSqueezingPainter}, i.e. $\eta=1$, is preferable for our aim to create entanglement. In this regime, the reflectivity of the coherent driving fields are almost unitary. Therefore, the reflected beams include many photons implying bright output beams. The violation of inequality Eq. (\ref{eq:FullV}) implies the entanglement of strong CV Gaussian states. Although one-mode squeezing of light has been demonstrated in optomechanical systems in both the unresolved-sideband regime \cite{NatureOneModeSqueezingPainter} and the resolved-sideband regime \cite{PRXOneModeSqueezing}, we will focus on the unresolved-sideband regime and the perfect coupling regime in our discussion below. We also take values for parameters, $\Delta=0, \Omega_{\text{m}0}=0.1\kappa, Q_m=10^6,\bar{n}_\text{th}=10^3,\kappa_\text{i}=0$.  \newline

\noindent{\bf Entanglement of two CV modes.} When two cavity modes are identical in detuning, $\Delta_j=\Delta=0$, optomechanical coupling strength, $G_j=G$, and decay rates, $\kappa_{\text{e},j}=\kappa_\text{e}$ and $\kappa_{\text{i},j}=\kappa_\text{i}$, the variances $V_{jl}$ are equal for any $j$ and $l$. If we drive the cavity modes properly, we are able to create entanglement of multicolor CVs. For example, the variance can be below the Duan's bound (blue region in Fig. \ref{fig:system}(b)) in the centre of the reflected CV modes corresponding to the driving modes in the output, $|\omega - \Omega_m|<0.95 \Omega_m$ for $\theta=-\pi/4, -3\pi/4$. Entanglement also can be obtained over the frequency region of $|\omega - \Omega_m|>1.03 \Omega_m$ for $\theta=\pi/4, 3\pi/4$. The minimal variance can be down to almost $1$ at $\theta=\pi/2$, implying a violation of Duan's bound by $3~dB$. \newline

\noindent{\bf Four-color entanglement.} We now examine how the random varying optomechanical coupling and decay rates, which are hard to control precisely in experiments, affect the degree of entanglement. 
\begin{figure}
 \centering
 \includegraphics[width=0.9\linewidth]{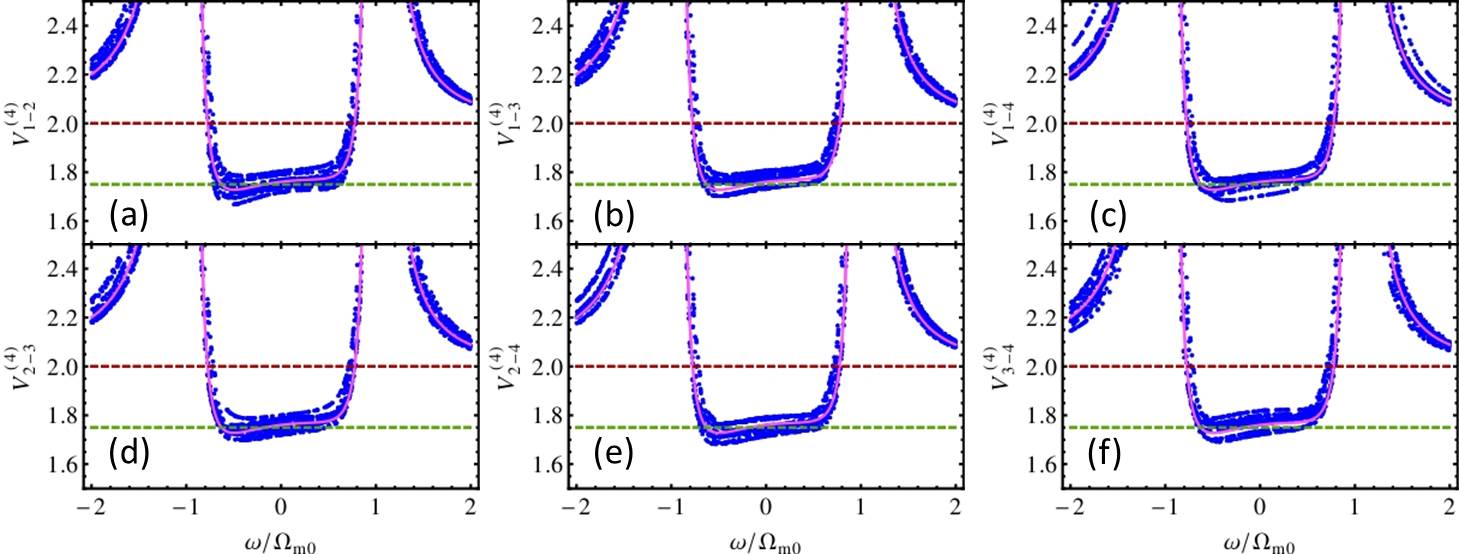}\\
 \caption{Four-mode entanglement evaluated over ten random samples. Parameters ($G_j$ and $\kappa_{e,j}$) of samples vary within $10\%$ of mean value with a normal distribution. Pink line indicates the variance for samples with identical cavity modes. Dashed red (green) line shows the Duan bound \cite{DuanCriterion1,DuanCriterion2} (the limit given by $V_\text{min}$). $G=0.5\Omega_{\text{m}0}, \theta=\pi/4$.} \label{fig:4MEntUni}
\end{figure}
In Fig. \ref{fig:4MEntUni}, we calculated all variance $V_{jl}$ for $j<l$ for four output CV modes. Pink lines show the variance for identical parameters and samples using Eq. (\ref{eq:FullCorrel}) in Eq. (\ref{eq:FullV}). To analyzing the influence of random variation of parameters in various samples, we present the variances of ten samples with independent normally distributed random optomechanical coupling rate, $G_j$ and decay rate, $\kappa_{\text{e},j}$ with variance of $10\%$ of their mean value. It is found that all six variances ($j<l$) can be as low as $1.75$, yielding a considerable four-color entanglement. \newline

\begin{figure}
 \centering
 \includegraphics[width=0.9\linewidth]{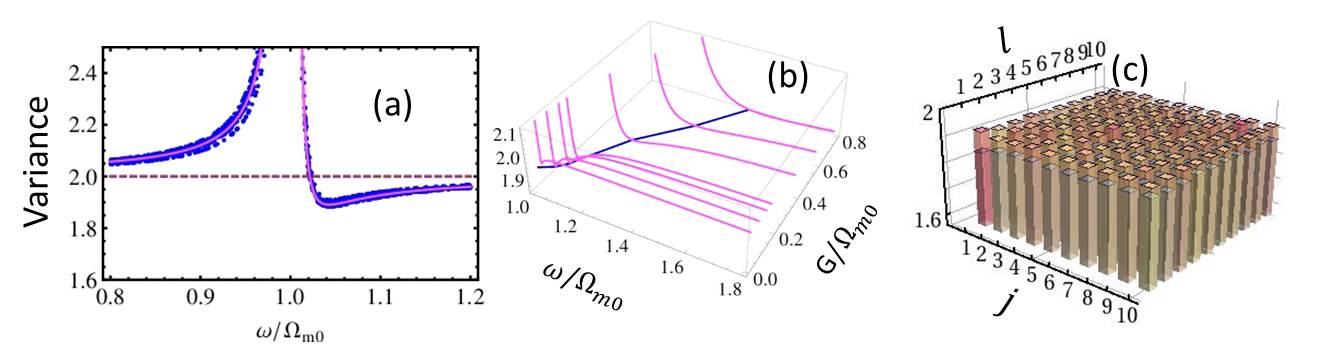}\\
 \caption{Entangled ten-mode CV cluster states. (a) Variance of $V^{(10)}_{1-2}$. Parameters of ten samples, $G_j$ and $\kappa_{\text{e},j}$, take a normal distribution with variance of $5\%$ of mean value. (b) Variance of $V^{(10)}_{jl}$ ($j\neq l$) (pink lines) of one sample with identical parameters, $G_j$ and $\kappa_{\text{e},j}$, as a function of $\omega$ for various coupling rates $G=\{0.02,0.05,0.08,0.1,0.2,0.3,0.4\}\times \Omega_{\text{m}0}$. Blue line shows the positions of smallest variances. (c) Variance matrix, $V^{(10)}_{jl}$ for $j,l\in\{1,2,\cdots,10\}$ at $\omega=1.04\Omega_{\text{m}0}$. Bars show the minimal and maximal variances evaluated over one hundred ($S=100$) samples with random coupling rate, $G_{m,j}$, and external coupling rate, $\kappa_{\text{e},m,j}$, varying within a normal distribution with variance of $\sigma=5\%$. $G=0.1\Omega_{\text{m}0}, \theta=-\pi/4$.} \label{fig:10MEnt}
\end{figure}

\noindent{\bf Ten-color entanglement.} Ten-color entangled CV cluster state is illustrated in Fig. \ref{fig:10MEnt}. Figure \ref{fig:10MEnt}(a) shows the variance evaluated by Eq. (\ref{eq:FullV}) using Eq. (\ref{eq:FullCorrel}) between modes $1$ and $2$ among ten CV modes. The pink line shows the variance $V_{12}^{(10)}$ for identical cavity modes. In the case of identical cavity modes all variance $V_{12}^{(jl)}$ with $j\neq l$ are equal. The blue spots show the variance $V_{12}^{(10)}$ of ten random samples with random parameters $G_j$ and $\kappa_{\text{e},j}$ taking a normal distribution with variance of $5\%$ of mean value. Obviously, this amount of variance in parameters only changes the entanglement nature slightly. The variance between any two of ten cavity modes in an optomechanical resonator with identical $G_j$ and $\kappa_{\text{e},j}$ for one sample is shown in Fig. \ref{fig:10MEnt}(b). The minimal variance $V_\text{min}$ is dependent on the optomechanical coupling, $G$. It decreases from $2$ to a limit of $2-\eta/M$ as $G$ increases, see blue line. Considering that the characteristic of solid-state devices is diverse from sample to sample, we examine the variance of each CV mode pairs over one hundred random samples, of which the parameters have the same mean value, but is distributed normally with variances of $5\%$ (see Fig. \ref{fig:10MEnt}(c)). Clearly, all variances, $V_{jl}^{(10)}$, are below the Duan bound,  although they may vary over a small range. The average is about $1.9$. It implies a ten-color entangled CV cluster state.
\newline

\noindent{\bf Experimental implementation.} We take a Fabry-P\'erot cavity containing a $\text{Si}_3\text{N}_4$ membrane in the middle as our optomechanical implementation. The whole system is at low temperature, $T\sim 500 \milli\kelvin$ \cite{ExpCavity1,ExpMR2}. For simplicity we assume that this FP cavity supports $10$ modes with different resonant frequencies but identical decay rates $\kappa/2\pi =1 \mega\hertz$ \cite{ExpCavity1} and $\kappa_i/2\pi$ negligible \cite{ExpCavity1}. The membrane we use has $Q_m=4\times 10^6$ and $\Omega_{\text{m}0}/2\pi =134  \kilo\hertz$ \cite{ExpMR2,ExpMR1} corresponding to the thermal phonon number of $\bar{n}_\text{th}=8\times 10^3$. We drive the cavity modes with a frequency comb from on-chip optomechanical resonators \cite{OptomOFC1,OptomOFC2}. Under on-resonance driving at each mode with input powers of $P_\text{in}=80 \micro\watt$ exciting the photon number of $\sim 1.4\times 10^8$ in each cavity mode around $\lambda\sim 1064 \nano\meter$. Such driving is strong enough to amplify optomechanical couplings to $0.1\Omega_{\text{m}0}$. In this optomechanical system the minimal variances of $V_\text{min}^{(10)}\sim 1.89<2$ is obtained at $\omega - \omega_{\text{L}j}\approx 1.052 \Omega_{\text{m}0}$ ($j\in \{1,2,\cdots,10\}$). Therefore, we create an entangled ten-color CV cluster state. If we apply our scheme to the microwave-optomechanical system developed by Andrews et al. \cite{MWOInterface2}, we can create entanglement between microwave and optical signals by reducing the variance to $V_\text{min}^{(2)}\sim 1.3$ ($\Omega_{\text{m}0}/2\pi=380 \kilo\hertz, Q_\text{m}>10^7,  \Omega_{\text{m}0}/\kappa \sim 0.25, T=40 \milli\kelvin, \bar{n}_\text{th}=2.2\times 10^3$ and $G=0.1\Omega_{\text{m}0}$). \newline 

\noindent{\bf Discussion} \\
\noindent In summary, we studied the multicolor entanglement of bright CV states from optomechanical systems. Entanglement up to ten CV modes has been demonstrated, while the output fields have many photons. It is found that the largest violation of Duan's bound among two-color CV modes is bounded by a limit of $3 \text{dB}$ ($|\theta|\sim \pi/2$). To create multicolor entanglement our on-chip optomechanical setup only requires on-resonance driving of the corresponding multimode of cavity. Our proposal does not rely on the probabilistic wave mixing process and therefore can also be extended to generate entanglement between microwave and optical photons \cite{MWOInterface1,MWOInterface2} or even in microwave frequency comb \cite{MWFC1,MWFC2}. Our work provides routes towards microwave or optical quantum frequency comb. It may enable quantum frequency comb-based applications in quantum information processing and quantum network. \newline

\noindent{\bf Acknowledgements} \\
K.X. acknowledge the support from the Australian Research
Council Centre of Excellence for Engineered Quantum Systems
(EQuS) (project number CE110001013) and the National Natural
Sciences Foundation of China (Grant No. 11204080). \newline


\FloatBarrier

%

%
%

%
%
%

\begin{center}
 \large{Supplementary material: Deterministic generation of bright multicolor entanglement from optomechanical systems}
\end{center}

%
%
%

\vskip 0.5cm


\nopagebreak

In this supplementary material we present the detailed derivation of the output noise power spectral density of continuous variables (CVs) and the formula for their variance between two different cavity modes.

\section{Model}

\subsection{Hamiltonian and quantum Langevin equations}
Our setup for multicolor entanglement of CVs is schematically illustrated in Figure 1(a). An optical frequency comb (OFC) laser incidents on the first optical grating and then is decomposited into a set of spectral lines with carrier frequency $\omega_{\text{L},j}$. Each spectral line is modulated in amplitude and phase by the liquid crystal light modulator for proper driving of cavity modes. These spectrally modulated OFC is combined by the second optical grating and then is applied to drive the cavity modes with rates $\kappa_{\text{e},j}$. The OFC field is reflected off the cavity to the output. The quantum noise in the output fields is squeezed due to the optomechanical interaction.

The Hamiltonian governing the evolution of multimode optomechanical system in Fig. 1 takes
\begin{equation}
  H  =\Omega_{\text{m}0} b^\dag b + \sum_j \Delta^\prime_j a_j^\dag a_j + \sum_j g_{\text{om},j}a_j^\dag a_j (b^\dag + b) 
 + i\sum_j \sqrt{2\kappa_{\text{e},j}} (\varepsilon_j a^\dag_j - \varepsilon_j^* a_j) \;,
\end{equation}
where $\Delta^\prime_j = \omega_j -\omega_{\text{L},j}$ is the detuning between the $j$th cavity mode and its driving, $g_{\text{om},j}$ is the optomechanical coupling between the cavity mode $a_j$ and the mechanical motion $b$.

The quantum Langevin equation reads \cite{QuntNoise1,QuntNoise2}
\begin{equation}
 \begin{split}
  \dot{a}_j & = -(i\Delta^\prime_j + \kappa_j)a_j - i g_{\text{om},j}(b^\dag + b) + \sqrt{2\kappa_{\text{e},j}} \varepsilon_j + \sqrt{2\kappa_{\text{e},j}}a_{\text{in,e},j}(t)  + \sqrt{2\kappa_{\text{i},j}}a_{\text{in,i},j}(t) \\
  \dot{b} & = -(i\Omega_{m0}+\gamma_{m0})b - i\sum_j g_{\text{om},j} a_j^\dag a_j + \sqrt{2\gamma_{\text{m}0}} b_\text{in}(t) \;,
 \end{split}
\end{equation}
Where $\kappa_{\text{i},j}$ and $\gamma_{\text{m}0}$ are the intrinsic loss rate of the $j$th cavity mode and the mechanical decay rate of mechanical resonator. All loss rates are necessarily accompanied by random fluctuating inputs $a_{\text{in,e},j}$, $a_{\text{in,i},j}$, and $b_\text{in}(t)$, for optical quantum noise from the external coupling channel, intrinsic loss channel and mechanical noise. We have $\langle a_{\text{in,e},j}\rangle=\langle a_{\text{in,i},j}\rangle=0$ and $\langle b_\text{in}\rangle=0$. 

In the steady state, we have
\begin{subequations}
\begin{align}
 \beta & = \langle b\rangle_{ss} = - \frac{i\sum_j g_{\text{om},j} |\bar{a}_j|^2}{i\Omega_{m0}+ \gamma_{m0}} \approx - \frac{1}{\Omega_{m0}}\sum_j g_{\text{om},j} |\bar{a}_j|^2 \;,\\
 \bar{a}_j & = \langle a_j\rangle_{ss} = \frac{\sqrt{2\kappa_{\text{e},j}}\varepsilon_j}{i\Delta_j + \kappa_j} \;,
\end{align}
\end{subequations}
with $\Delta_j = \Delta^\prime_j + 2 g_{\text{om},j} \beta$.

Now we linearize the operators as $a_j \rightarrow \bar{a}_j + a_j$ and $b \rightarrow \beta + b$ \cite{PhysRevLett.99.093901,PhysRevLett.99.093902}, and then have the Hamiltonian after the linearization reads as
\begin{equation}
 H^\prime = \Omega_{\text{m}0} b^\dag b + \sum_j \Delta_j a^\dag_j a_j + \sum_j (G_j a^\dag_j + G_j^* a_j)(b^\dag + b) \;,
\end{equation}
with $G_j=g_{\text{om},j} \bar{a}_j$.
Then the quantum Langevin equations (QLEs) for $a$ and $b$ becomes
\begin{subequations}
\begin{align}
 \dot{a}_j & = -(i\Delta_j + \kappa_j)a_j - i G_j(b^\dag + b)  + \sqrt{2\kappa_{e,j}}a_{\text{in,e},j}  + \sqrt{2\kappa_{i,j}}a_{\text{in,i},j} \\
  \dot{b} & = -(i\Omega_{m0}+\gamma_{m0})b - i\sum_j  (G_ja_j^\dag +G^*_ja_j) + \sqrt{2\gamma_{\text{m}0}} b_\text{in}(t) \;,
\end{align}
\end{subequations}
with the effective detuning becomes $\Delta_j = \omega_j - \omega_{L,j} + 2 g_{om,j} \beta$ and the total decay rate of cavity mode is $\kappa_j = \kappa_{e,j} + \kappa_{i,j}$. Below we assume all $G_j$ are real because we can tune the detuning and the driving $\varepsilon_j$. 

To solve the Langevin equation in frequency domain, we define the Fourier transformation
\begin{subequations}
 \begin{align}
  \hat{A}(t) & = \frac{1}{\sqrt{2\pi}} \int_{-\infty}^\infty d\omega e^{-i\omega t} \hat{A}(\omega) \;,\\
  \hat{A}(\omega) & =\frac{1}{\sqrt{2\pi}} \int_{-\infty}^\infty d\omega e^{i\omega t} \hat{A}(t) \;.
 \end{align}
\end{subequations}
The optical cavity susceptibility for the $j$th cavity mode is 
\begin{equation}
 \chi_{R,j}(\omega) = \frac{1}{i(\Delta_j - \omega) + \kappa_j} \;.
\end{equation}
The mechanical susceptibility $\chi_\text{m}(\omega)$, 
\begin{equation}\label{eq:chim}
\chi^{-1}_m(\omega) = i (\Omega_{m0} -\omega) + \gamma_{m0} + \sum_j G_j^{2}\left[  \chi_{R,j}(\omega) - \chi^*_{R,j}(-\omega) \right] \;,
\end{equation}
connects the mechanical response to the environmental noise and the optical vacuum fluctuations incident on the optical cavity.
Applying the relation $[A(-\omega)]^\dag=A^\dag(\omega)$ and $[A^\dag(-\omega)]^\dag=A(\omega)$, we have \cite{NatureOneModeSqueezingPainter}
\begin{subequations}\label{eq:abomega}
 \begin{align}
  a_j(\omega) & = \sqrt{2\kappa_{\text{e},j}} \chi_{R,j}(\omega) a_{\text{in,e},j}(\omega) + \sqrt{2\kappa_{\text{i},j}} \chi_{R,j}(\omega) a_{\text{in,i},j}(\omega) -iG_j \chi_{R,j}(\omega) \left[b^\dag(\omega)  + b(\omega)\right] \;, \\
  b(\omega) & = \frac{\sqrt{2\gamma_{\text{m}0}} b_\text{in}(\omega) -i\sum_j G_j\left[ a^\dag_j(\omega) + a_j(\omega) \right]}{i(\Omega_{\text{m}0}-\omega) + \gamma_{\text{m}0}} \;.
 \end{align}
\end{subequations}
At temperature of $T$, the correlations for noises are $\langle a_{\text{in,e},j}(\omega)a^\dag_{\text{in,e},l}(\omega^\prime) \rangle=(\bar{n}_{\text{o},j}+1)\delta_{jl} \delta(\omega+\omega^\prime)$, $\langle a_{\text{in,i},j}(\omega)a^\dag_{\text{in,i},l}(\omega^\prime) \rangle=(\bar{n}_{\text{o},j}+1)\delta_{jl} \delta(\omega+\omega^\prime)$, where $\delta_{jl}$ ($\delta(\omega+\omega^\prime)$) is the discrete Kronecker (Dirac) delta function and $\bar{n}_{\text{o},j} = (e^{\hbar \omega_j/K_BT}-1)^{-1}$. Typically, $\bar{n}_{\text{o},j} \approx 0$ at light frequency. We also have the correlation for mechanical noise, $\langle b_\text{in}(\omega)b^\dag_\text{in}(\omega^\prime)= (\bar{n}_\text{th}+1)\delta(\omega+\omega^\prime)$ and $\langle b^\dag_\text{in}(\omega)b_\text{in}(\omega^\prime)= \bar{n}_\text{th}\delta(\omega+\omega^\prime)$ with $\bar{n}_\text{th} = (e^{\hbar \Omega_{\text{m}0}/K_BT}-1)^{-1}$.
$\bar{n}_{\text{o},j}$ and $\bar{n}_\text{th}$ are the thermal photon and phonon occupancies, respectively.

By manipulation of Eq. (\ref{eq:abomega}) the mechanical motion in frequency domain can be expressed as the response to mechanical noise and the optical vacuum fluctuations through the optomechanical coupling
\begin{subequations} \label{eq:bomega}
 \begin{align}
  b(\omega) & = \sqrt{2\gamma_{\text{m}0}} \chi_\text{m}(\omega) b_\text{in}(\omega) - i\chi_\text{m}(\omega)\sum_jG_j \sqrt{2\kappa_{\text{e},j}} \left[\chi_{R,j}(\omega)a_{\text{in,e},j}(\omega) +  \chi^*_{R,j}(-\omega)a^\dag_{\text{in,e},j}(\omega) \right] \\ \nonumber
  & - i\chi_\text{m}(\omega)\sum_jG_j \sqrt{2\kappa_{\text{i},j}} \left[\chi_{R,j}(\omega) a_{\text{in,i},j}(\omega) +  \chi^*_{R,j}(-\omega)a^\dag_{\text{in,i},j}(\omega) \right]
 \;. 
 \end{align}
\end{subequations}

Substituting Eq. \ref{eq:bomega} into $a_j(\omega)$, we have
\begin{equation} \label{eq:aomega}
 \begin{split}
  a_j(\omega) & = \sqrt{2\kappa_{e,j}}\chi_{R,j}(\omega)a_{in,e,j}(\omega) + \sqrt{2\kappa_{ij}}\chi_{Rj}(\omega) a_{in,i,j}(\omega)\\
 & -i\sqrt{2\gamma_{m0}} G_j\chi_{Rj}(\omega) \left[\chi_m(\omega)b_{in}(\omega) + \chi_m^*(-\omega)b^\dag_{in}(\omega) \right] \\
 & + G_j \chi_{Rj}(\omega) [\chi^*_m(-\omega) - \chi_m(\omega)]\sum_n G_n \sqrt{2\kappa_{en}} [\chi_{Rn}(\omega) a_{in,e,n}(\omega) + \chi^*_{Rn}(-\omega) a^\dag_{in,e,n}(\omega)]\\
& + G_j \chi_{Rj}(\omega) [\chi^*_m(-\omega) - \chi_m(\omega)]\sum_n G_n \sqrt{2\kappa_{in}} [\chi_{Rn}(\omega) a_{in,i,n}(\omega) + \chi^*_{Rn}(-\omega) a^\dag_{in,i,n}(\omega)] \;.
 \end{split}
\end{equation}
Using this equation we are able to calculate the power spectal density and the variance of the output noise.

\subsection{Output field}
The output field connects to the cavity mode field through the input-output relation, $a_{\text{out},j}(\omega) = - a_{\text{in,e},j}(\omega) + \sqrt{2\kappa_{e,j}} a_j(\omega)$. Defining $F_\text{m}(\omega) = \chi^*_m(-\omega) - \chi_m(\omega)$ (Note that $F_m^*(-\omega)=-F_\text{m}(\omega)$), then we have the output field
\begin{equation}
 \begin{split}
   a_{\text{out},j}(\omega) & = [-1+2\kappa_{e,j}\chi_{R,j}(\omega)]a_{in,e,j}(\omega) + 2\sqrt{\kappa_{ej}\kappa_{ij}}\chi_{Rj}(\omega) a_{in,i,j}(\omega)\\
 & -i\sqrt{2\gamma_{m0}} \sqrt{2\kappa_{\text{e},j}}G_j\chi_{Rj}(\omega) \left[\chi_m(\omega)b_{in}(\omega) + \chi_m^*(-\omega)b^\dag_{in}(\omega) \right] \\
 & + 2\kappa_{ej} G_j \chi_{Rj}(\omega) F_m(\omega)\sum_n G_n \sqrt{\frac{\kappa_{en}}{\kappa_{ej}}} [\chi_{Rn}(\omega) a_{in,e,n}(\omega) + \chi^*_{Rn}(-\omega) a^\dag_{in,e,n}(\omega)]\\
& + 2\kappa_{ej} G_j \chi_{Rj}(\omega)F_m(\omega) \sum_n G_n \sqrt{\frac{\kappa_{in}}{\kappa_{ej}}} [\chi_{Rn}(\omega) a_{in,i,n}(\omega) + \chi^*_{Rn}(-\omega) a^\dag_{in,i,n}(\omega)] \;.
 \end{split}
\end{equation}

To study the entanglement of the output field, we define the fluctuation operators for $X$ and $Y$ quadratures as \cite{NatureOneModeSqueezingPainter,ThreeColorEnt1}
\begin{subequations} \label{eq:XYQuadrature}
\begin{align}
 & \Delta X_{\text{out},j}(\omega)  = \frac{e^{-i\theta} a_{\text{out},j}(\omega) + e^{i\theta} a^\dag_{\text{out},j}(\omega)}{\sqrt{2}} \;, \\
 & \Delta Y_{\text{out},j}(\omega)  = \frac{e^{-i\theta} a_{\text{out},j}(\omega) - e^{i\theta} a^\dag_{\text{out},j}(\omega)}{i\sqrt{2}} \;.
\end{align}
\end{subequations}


\subsection{X quadrature}
We define $X$ quadrature of the output as
\begin{equation} 
\begin{split}
 & \Delta X_{\text{out},j}(\omega)  = \frac{e^{-i\theta} a_{\text{out},j}(\omega) + e^{i\theta} a^\dag_{\text{out},j}(\omega)}{\sqrt{2}}\\
 & =\frac{1}{\sqrt{2}}e^{-i\theta}[2\kappa_{e,j}\chi_{R,j}(\omega)-1]a_{\text{in,e},j}(\omega) + \frac{1}{\sqrt{2}}e^{i\theta}[2\kappa_{e,j}\chi^*_{R,j}(-\omega)-1]a^\dag_{\text{in,e},j}(\omega) \\
 & + \sqrt{2\kappa_{ej}\kappa_{ij}}e^{-i\theta}\chi_{Rj}(\omega) a_{in,i,j}(\omega)+ \sqrt{2\kappa_{ej}\kappa_{ij}}e^{i\theta}\chi^*_{Rj}(-\omega) a^\dag_{in,i,j}(\omega)  \\
& +i \sqrt{2\kappa_{\text{e},j}}\sqrt{\gamma_{m0}}G_j \zeta_j(\omega) \left[\chi_m(\omega)b_{in}(\omega) + \chi_m^*(-\omega)b^\dag_{in}(\omega) \right] \\
& -\sqrt{2}\kappa_{ej} G_jF_m(\omega) \zeta_j(\omega) \sum_n G_n \sqrt{\frac{\kappa_{en}}{\kappa_{ej}}} [\chi_{Rn}(\omega) a_{in,e,n}(\omega) + \chi^*_{Rn}(-\omega) a^\dag_{in,e,n}(\omega)]  \\
& -\sqrt{2}\kappa_{ej} G_jF_m(\omega)\zeta_j(\omega)  \sum_n G_n \sqrt{\frac{\kappa_{in}}{\kappa_{ej}}} [\chi_{Rn}(\omega) a_{in,i,n}(\omega) + \chi^*_{Rn}(-\omega) a^\dag_{in,i,n}(\omega)] \;,
\end{split}
\end{equation}
with $\zeta_j(\omega) = \left[ e^{i\theta}\chi^*_{Rj}(-\omega) -e^{-i\theta}\chi_{Rj}(\omega) \right]$ and $ \zeta_j(-\omega)=- \zeta^*_j(\omega)$.

Thus, the correlation of $x$ quadrature between the $j$th and $l$th cavity modes can be evaluated by
\begin{equation}
 \begin{split}
 & \langle \Delta X_{\text{out},j}(\omega) \Delta X_{\text{out},l}(-\omega)\rangle  = 2\gamma_{m0}\sqrt{\kappa_{ej}\kappa_{el}} G_j G_l \zeta_j(\omega)\zeta^*_l(\omega)\left[|\chi_m(\omega)|^2 (\bar{n}_m +1) + |\chi_m(-\omega)|^2 \bar{n}_m\right] \\
 & + \frac{1}{2} |2\kappa_{e,j}\chi_{R,j}(\omega)-1|^2 (\bar{n}_{o,j} +1) \delta_{jl} + \frac{1}{2} |2\kappa_{e,j}\chi_{R,j}(-\omega)-1|^2 \bar{n}_{o,j} \delta_{jl} \\
& - e^{-i\theta}\sqrt{\kappa_{ej}\kappa_{el}} G_lG_j \left[2\kappa_{e,j}\chi_{R,j}(\omega)-1 \right] F_m^*(\omega) \zeta^*_l(\omega)\chi^*_{R,j}(\omega) (\bar{n}_{o,j} +1) \\
& - e^{i\theta}\sqrt{\kappa_{ej}\kappa_{el}} G_lG_j \left[2\kappa_{e,j}\chi^*_{R,j}(-\omega)-1 \right] F_m^*(\omega) \zeta^*_l(\omega)\chi_{R,j}(-\omega) \bar{n}_{o,j}  \\
& - e^{i\theta}\sqrt{\kappa_{ej}\kappa_{el}} G_lG_j \left[2\kappa_{e,l}\chi^*_{R,l}(\omega)-1 \right] F_m(\omega) \zeta_j(\omega)\chi_{R,l}(\omega) (\bar{n}_{o,l} +1) \\
& - e^{-i\theta}\sqrt{\kappa_{ej}\kappa_{el}} G_lG_j \left[2\kappa_{e,l}\chi_{R,l}(-\omega)-1 \right] F_m(\omega) \zeta_l(\omega)\chi^*_{R,l}(-\omega) \bar{n}_{o,l}\\
& + 2 \kappa_{e,j} \kappa_{i,j} |\chi_{R,j}(\omega)|^2 (\bar{n}_{o,j} +1) \delta_{jl}   + 2 \kappa_{e,j} \kappa_{i,j} |\chi_{R,j}(-\omega)|^2 \bar{n}_{o,j} \delta_{jl}  \\ 
& - 2\kappa_{i,j} \sqrt{\kappa_{e,l}\kappa_{e,j}} e^{-i\theta}G_l G_j |\chi_{R,j}(\omega)|^2F_m^*(\omega) \zeta^*_l(\omega)(\bar{n}_{o,j} +1)  \\
& - 2 \kappa_{i,j} \sqrt{\kappa_{e,l}\kappa_{e,j}} e^{i\theta}G_l G_j|\chi_{R,j}(-\omega)|^2F_m^*(\omega) \zeta^*_l(\omega)\bar{n}_{o,j}  \\
& - 2\kappa_{i,l} \sqrt{\kappa_{e,l}\kappa_{e,j}} e^{i\theta}G_l G_j |\chi_{R,l}(\omega)|^2F_m(\omega) \zeta_j(\omega)(\bar{n}_{o,l} +1)  \\
& - 2 \kappa_{i,l} \sqrt{\kappa_{e,l}\kappa_{e,j}} e^{-i\theta}G_l G_j|\chi_{R,l}(-\omega)|^2F_m(\omega) \zeta_j(\omega)\bar{n}_{o,l}  \\
& + 2 \sqrt{\kappa_{e,j}\kappa_{e,l}} G_l G_j |F_m(\omega)|^2 \zeta_j(\omega)\zeta^*_l(\omega)\sum_n (\kappa_{e,n} + \kappa_{i,n}) G_n^2 |\chi_{R,n}(\omega)|^2 (\bar{n}_{o,n}+1)\\
& + 2 \sqrt{\kappa_{e,j}\kappa_{e,l}} G_l G_j |F_m(\omega)|^2 \zeta_j(\omega)\zeta^*_l(\omega)\sum_n (\kappa_{e,n} + \kappa_{i,n}) G_n^2 |\chi_{R,n}(-\omega)|^2 \bar{n}_{o,n}\;.
\end{split}
\end{equation}

Merging some terms, we have
\begin{equation}
 \begin{split}
 & \langle \Delta X_{\text{out},j}(\omega) \Delta X_{\text{out},l}(-\omega)\rangle  =\frac{1}{2} (2\bar{n}_{\text{o},j}+1) \delta_{jl} \\
& + 2\gamma_{m0}\sqrt{\kappa_{ej}\kappa_{el}} G_j G_l \zeta_j(\omega)\zeta^*_l(\omega)\left[|\chi_m(\omega)|^2 (\bar{n}_m +1) + |\chi_m(-\omega)|^2 \bar{n}_m\right] \\
& - e^{-i\theta}\sqrt{\kappa_{ej}\kappa_{el}} G_lG_j \left[2\kappa_j\chi_{R,j}(\omega)-1 \right] F_m^*(\omega) \zeta^*_l(\omega)\chi^*_{R,j}(\omega) (\bar{n}_{\text{o},j}+1) \\
& - e^{i\theta}\sqrt{\kappa_{ej}\kappa_{el}} G_lG_j \left[2\kappa_l\chi^*_{R,l}(\omega)-1 \right] F_m(\omega) \zeta_j(\omega)\chi_{R,l}(\omega) (\bar{n}_{\text{o},l}+1)\\
& + 2 \sqrt{\kappa_{e,j}\kappa_{e,l}} G_l G_j |F_m(\omega)|^2 \zeta_j(\omega)\zeta^*_l(\omega)\sum_n (\kappa_{e,n} + \kappa_{i,n}) G_n^2 |\chi_{R,n}(\omega)|^2  (\bar{n}_{\text{o},n}+1) \\
& - e^{-i\theta}\sqrt{\kappa_{ej}\kappa_{el}} G_lG_j \left[2\kappa_j\chi^*_{R,j}(-\omega)-1 \right] F_m^*(\omega) \zeta^*_l(\omega)\chi_{R,j}(-\omega) \bar{n}_{\text{o},j} \\
& - e^{i\theta}\sqrt{\kappa_{ej}\kappa_{el}} G_lG_j \left[2\kappa_l\chi_{R,l}(-\omega)-1 \right] F_m(\omega) \zeta_j(\omega)\chi^*_{R,l}(-\omega) \bar{n}_{\text{o},l}\\
& + 2 \sqrt{\kappa_{e,j}\kappa_{e,l}} G_l G_j |F_m(\omega)|^2 \zeta_j(\omega)\zeta^*_l(\omega)\sum_n \kappa_{n}  G_n^2 |\chi_{R,n}(-\omega)|^2  \bar{n}_{\text{o},n} \;.
\end{split}
\end{equation}

\subsection{Y quadrature}

Similar to $X$ quadrature, we define $Y$ quadrature as
\begin{equation} 
\begin{split}
 & \Delta Y_{\text{out},j}(\omega)  = \frac{e^{-i\theta} a_{\text{out},j}(\omega) - e^{i\theta} a^\dag_{\text{out},j}(\omega)}{i\sqrt{2}}\\
 & =\frac{1}{i\sqrt{2}}e^{-i\theta}[2\kappa_{e,j}\chi_{R,j}(\omega)-1]a_{\text{in,e},j}(\omega) - \frac{1}{i\sqrt{2}}e^{i\theta}[2\kappa_{e,j}\chi^*_{R,j}(-\omega)-1]a^\dag_{\text{in,e},j}(\omega)  \\
 & -i \sqrt{2\kappa_{ej}\kappa_{ij}}e^{-i\theta}\chi_{Rj}(\omega) a_{in,i,j}(\omega)+ i\sqrt{2\kappa_{ej}\kappa_{ij}}e^{i\theta}\chi^*_{Rj}(-\omega) a^\dag_{in,i,j}(\omega) \\
& - \sqrt{2\kappa_{\text{e},j}}\sqrt{\gamma_{m0}}G_j \varXi_j(\omega) \left[\chi_m(\omega)b_{in}(\omega) + \chi_m^*(-\omega)b^\dag_{in}(\omega) \right]  \\
& -i\sqrt{2}\kappa_{ej} G_jF_m(\omega) \varXi_j(\omega) \sum_n G_n \sqrt{\frac{\kappa_{en}}{\kappa_{ej}}} [\chi_{Rn}(\omega) a_{in,e,n}(\omega) + \chi^*_{Rn}(-\omega) a^\dag_{in,e,n}(\omega)]  \\
& -i\sqrt{2}\kappa_{ej} G_jF_m(\omega)\varXi_j(\omega)  \sum_n G_n \sqrt{\frac{\kappa_{in}}{\kappa_{ej}}} [\chi_{Rn}(\omega) a_{in,i,n}(\omega) + \chi^*_{Rn}(-\omega) a^\dag_{in,i,n}(\omega)]\;,
\end{split}
\end{equation}
with $\varXi_j(\omega) = \left[ e^{i\theta}\chi^*_{Rj}(-\omega) +e^{-i\theta}\chi_{Rj}(\omega) \right]$ and $\varXi^*_j(-\omega) = \varXi_j(\omega)$, and have the correlation
%
\begin{equation}
 \begin{split}
 & \langle \Delta Y_{\text{out},j}(\omega) \Delta Y_{\text{out},l}(-\omega)\rangle  = 2\gamma_{m0}\sqrt{\kappa_{ej}\kappa_{el}} G_j G_l \varXi_j(\omega)\varXi^*_l(\omega)\left[|\chi_m(\omega)|^2 (\bar{n}_m +1) + |\chi_m(-\omega)|^2 \bar{n}_m\right]\\
 & + \frac{1}{2} |2\kappa_{e,j}\chi_{R,j}(\omega)-1|^2 (\bar{n}_{o,j} +1) \delta_{jl} + \frac{1}{2} |2\kappa_{e,j}\chi_{R,j}(-\omega)-1|^2 \bar{n}_{o,j} \delta_{jl}  \\
& + e^{-i\theta}\sqrt{\kappa_{ej}\kappa_{el}} G_lG_j \left[2\kappa_{e,j}\chi_{R,j}(\omega)-1 \right] F_m^*(\omega) \varXi^*_l(\omega)\chi^*_{R,j}(\omega) (\bar{n}_{o,j} +1) \\
& - e^{i\theta}\sqrt{\kappa_{ej}\kappa_{el}} G_lG_j \left[2\kappa_{e,j}\chi^*_{R,j}(-\omega)-1 \right] F_m^*(\omega) \varXi^*_l(\omega)\chi_{R,j}(-\omega) \bar{n}_{o,j}  \\
& + e^{i\theta}\sqrt{\kappa_{ej}\kappa_{el}} G_lG_j \left[2\kappa_{e,l}\chi^*_{R,l}(\omega)-1 \right] F_m(\omega) \varXi_j(\omega)\chi_{R,l}(\omega) (\bar{n}_{o,l} +1)  \\
& - e^{-i\theta}\sqrt{\kappa_{ej}\kappa_{el}} G_lG_j \left[2\kappa_{e,l}\chi_{R,l}(-\omega)-1 \right] F_m(\omega) \varXi_l(\omega)\chi^*_{R,l}(-\omega) \bar{n}_{o,l} \\
& + 2 \kappa_{e,j} \kappa_{i,j} |\chi_{R,j}(\omega)|^2 (\bar{n}_{o,j} +1) \delta_{jl}   + 2 \kappa_{e,j} \kappa_{i,j} |\chi_{R,j}(-\omega)|^2 \bar{n}_{o,j} \delta_{jl}  \\ 
& + 2\kappa_{i,j} \sqrt{\kappa_{e,l}\kappa_{e,j}} e^{-i\theta}G_l G_j |\chi_{R,j}(\omega)|^2F_m^*(\omega) \varXi^*_l(\omega)(\bar{n}_{o,j} +1)   \\
& - 2 \kappa_{i,j} \sqrt{\kappa_{e,l}\kappa_{e,j}} e^{i\theta}G_l G_j|\chi_{R,j}(-\omega)|^2F_m^*(\omega) \varXi^*_l(\omega)\bar{n}_{o,j} \\
& + 2\kappa_{i,l} \sqrt{\kappa_{e,l}\kappa_{e,j}} e^{i\theta}G_l G_j |\chi_{R,l}(\omega)|^2F_m(\omega) \varXi_j(\omega)(\bar{n}_{o,l} +1)   \\
& - 2 \kappa_{i,l} \sqrt{\kappa_{e,l}\kappa_{e,j}} e^{-i\theta}G_l G_j|\chi_{R,l}(-\omega)|^2F_m(\omega) \varXi_j(\omega)\bar{n}_{o,l}\\
& + 2 \sqrt{\kappa_{e,j}\kappa_{e,l}} G_l G_j |F_m(\omega)|^2 \varXi_j(\omega)\varXi^*_l(\omega)\sum_n (\kappa_{e,n} + \kappa_{i,n}) G_n^2 |\chi_{R,n}(\omega)|^2 (\bar{n}_{o,n}+1)\\
& + 2 \sqrt{\kappa_{e,j}\kappa_{e,l}} G_l G_j |F_m(\omega)|^2 \varXi_j(\omega)\varXi^*_l(\omega)\sum_n (\kappa_{e,n} + \kappa_{i,n}) G_n^2 |\chi_{R,n}(-\omega)|^2 \bar{n}_{o,n} \;.
\end{split}
\end{equation}

Merging some terms, we have
\begin{equation}
 \begin{split}
 & \langle \Delta Y_{\text{out},j}(\omega) \Delta Y_{\text{out},l}(-\omega)\rangle  =
 \frac{1}{2} (2\bar{n}_{\text{o},j}+1)  \delta_{jl} \\
 & +2\gamma_{m0}\sqrt{\kappa_{ej}\kappa_{el}} G_j G_l \varXi_j(\omega)\varXi^*_l(\omega)\left[|\chi_m(\omega)|^2 (\bar{n}_m +1) + |\chi_m(-\omega)|^2 \bar{n}_m\right]\\
& + e^{-i\theta}\sqrt{\kappa_{ej}\kappa_{el}} G_lG_j \left[2\kappa_j\chi_{R,j}(\omega)-1 \right] F_m^*(\omega) \varXi^*_l(\omega)\chi^*_{R,j}(\omega)  (\bar{n}_{\text{o},j}+1) \\
& + e^{i\theta}\sqrt{\kappa_{ej}\kappa_{el}} G_lG_j \left[2\kappa_l\chi^*_{R,l}(\omega)-1 \right] F_m(\omega) \varXi_j(\omega)\chi_{R,l}(\omega)  (\bar{n}_{\text{o},l}+1)  \\
& + 2 \sqrt{\kappa_{e,j}\kappa_{e,l}} G_l G_j |F_m(\omega)|^2 \varXi_j(\omega)\varXi^*_l(\omega)\sum_n (\kappa_{e,n} + \kappa_{i,n}) G_n^2 |\chi_{R,n}(\omega)|^2  (\bar{n}_{\text{o},n}+1)\\
%
& + e^{-i\theta}\sqrt{\kappa_{ej}\kappa_{el}} G_lG_j \left[2\kappa_j\chi^*_{R,j}(-\omega)-1 \right] F_m^*(\omega) \varXi^*_l(\omega)\chi_{R,j}(-\omega) \bar{n}_{\text{o},j}  \\
& + e^{i\theta}\sqrt{\kappa_{ej}\kappa_{el}} G_lG_j \left[2\kappa_l\chi_{R,l}(-\omega)-1 \right] F_m(\omega) \varXi_j(\omega)\chi^*_{R,l}(-\omega) \bar{n}_{\text{o},l}  \\
& + 2 \sqrt{\kappa_{e,j}\kappa_{e,l}} G_l G_j |F_m(\omega)|^2 \varXi_j(\omega)\varXi^*_l(\omega)\sum_n \kappa_{n} G_n^2 |\chi_{R,n}(-\omega)|^2\bar{n}_{\text{o},n} \;.
\end{split}
\end{equation}

\section{Entanglement}
One of the sufficient criterion for the bipartite entanglement of any two CV states is Duan's criteria \cite{DuanCriterion1}. Its extension for multipartite entanglement of CV states is written directly in terms of the correlation of fluctuation of fields, as sums of variances \cite{DuanCriterion2}. Here we derive the variances for any paired CV states to provide an analysis for multicolor entanglement of CVs.

We assume that there are $M$ cavity modes in our optomechanical system involved in operation. Without loss of generality, we calculate the sums of variances between two modes $a_j$ and $a_l$ by 
\begin{equation}
\begin{split}
 V^{(M)}_{jl}(\omega) = & (\Delta X_{\text{out},j}(\omega)- \Delta X_{\text{out},l}(\omega))(\Delta X_{\text{out},j}(-\omega)- \Delta X_{\text{out},l}(-\omega)) \\
 & +  (\Delta Y_{\text{out},j}(\omega)+\Delta Y_{\text{out},l}(\omega)) (\Delta Y_{\text{out},j}(-\omega)+\Delta Y_{\text{out},l}(-\omega)) \;.
\end{split}
\end{equation}
 The sufficient criterion for inseparability between these two modes is $V_{jl}<2$. If all bipartition in $M$ CV states are entangled, i.e. $V^{(M)}_{jl}<2$ for any $j\neq l \in \{1,2,\cdots,M\}$, genuine multipartite entanglement is then obtained \cite{ThreeColorEnt1,NatPhys11.167-SqueezingComb,EntComb}. Note that the way to calculate these variances is not optimal for the criteria for analyzing CV entanglement. But the violation of inequality $V_{jl}\geq 2$ is sufficient to claim that these two modes are entangled. This variance can be smaller if coefficients associated with other modes are properly chosen \cite{DuanCriterion2,ThreeColorEnt1}.

 To provide a fundamental limitation for obtainable entanglement we consider the case all cavity modes have identical parameters, i.e. $G_j=G, \kappa_{\text{e},j}=\kappa_\text{e},\kappa_{\text{i},j}=\kappa_\text{i},\kappa_j=\kappa$ and $\Delta_j=\Delta$. We take $\Delta \sim 0$ and define $\delta_\pm = \Omega_{m0}\pm\omega$ and $\delta^{-1} = \delta_-^{-1}+ \delta_+^{-1}$.
Assuming that $|\delta|\gg \gamma_{m0}$, then we have
\begin{enumerate}
 \item $\chi_\text{R}(\omega) =\frac{1}{-i\omega + \kappa}$;
 
 \item $\chi_m(\omega) \approx -\frac{i}{\delta_-}$ and $\chi_m(-\omega) \approx -\frac{i}{\delta_+}$;

 \item $F_m(\omega) \approx \frac{i}{\delta}$;

 \item $\zeta_j(\omega) = \frac{2i\sin\theta}{-i\omega + \kappa}$; $\zeta^*_j(\omega) = -\zeta_j(-\omega)$;

 \item $\varXi_j(\omega) = \frac{2\cos\theta}{-i\omega + \kappa}$ and $\varXi^*_j(\omega) =\varXi_j(-\omega)$.
\end{enumerate}

Using the above approximation we obtain the correlation
\begin{subequations} \label{eq:SimpCorrel}
 \begin{align}
  \langle \Delta X_{\text{out}(\omega),j}\Delta X_{\text{out},l}(-\omega)\rangle = &  \frac{1}{2} (2\bar{n}_\text{o} +1) \delta_{jl}  + \eta M\left(\frac{\tilde{\Gamma}_\text{meas}}{2\delta}\right)^2(1-\cos2\theta) (2\bar{n}_\text{o} +1) \\ \nonumber
 & + \eta \frac{\tilde{\Gamma}_\text{meas}}{\delta} \left[\frac{\Omega_m}{\delta}\frac{\bar{n}_m}{Q_m}\delta^2 (\delta_-^{-2}+\delta_+^{-2})(1-\cos2\theta) + \frac{\sin2\theta}{2} \right](2\bar{n}_\text{o} +1)\;,\\
  \langle \Delta Y_{\text{out}(\omega),j}\Delta Y_{\text{out},l}(-\omega)\rangle =&  \frac{1}{2}(2\bar{n}_\text{o} +1)\delta_{jl}  + \eta M\left(\frac{\tilde{\Gamma}_\text{meas}}{2\delta}\right)^2(1+\cos2\theta)(2\bar{n}_\text{o} +1) \\ \nonumber
 &  + \eta \frac{\tilde{\Gamma}_\text{meas}}{\delta} \left[\frac{\Omega_m}{\delta}\frac{\bar{n}_m}{Q_m}\delta^2 (\delta_-^{-2}+\delta_+^{-2})(1+\cos2\theta) - \frac{\sin2\theta}{2} \right](2\bar{n}_\text{o} +1)\;,
 \end{align}
\end{subequations}
with $\eta = \kappa_e/\kappa$ and $\tilde{\Gamma}_\text{meas}=\frac{4G^2}{\kappa} \frac{\kappa^2}{\omega^2+\kappa^2}$, and the variance 
\begin{equation} \label{eq:FullV}
\begin{split}
 V^{(M)}_{jl}(\omega) = & 2 (2\bar{n}_\text{o} +1) + 4\eta M\left(\frac{\tilde{\Gamma}_\text{meas}}{2\delta}\right)^2(1+\cos2\theta)(2\bar{n}_\text{o} +1) \\
 & + 4\eta \frac{\tilde{\Gamma}_\text{meas}}{\delta} \left[\frac{\Omega_m}{\delta}\frac{\bar{n}_m}{Q_m}\delta^2 (\delta_-^{-2}+\delta_+^{-2})(1+\cos2\theta) - \frac{\sin2\theta}{2} \right](2\bar{n}_\text{o} +1) \;.
\end{split}
\end{equation}
Obviously, our Eq. (\ref{eq:SimpCorrel}) for a single mode, $j=l$, agrees with the output noise power density in \cite{NatureOneModeSqueezingPainter} after replacing $\kappa$ with $\kappa/2$ and setting $\bar{n}_o=0$.
Entanglement of bipartite CV states is obtained when $|\delta| > \left(\frac{M\tilde{\Gamma}_\text{meas}}{2} + \frac{\Omega_\text{m}\bar{n}_m}{Q_m}\right) \left| \frac{\cos\theta}{\sin\theta}\right|(2\bar{n}_\text{o} +1)$. While the variance $V^{(M)}_{jl}$ is minimal, $V^{(M)}_\text{min}=2 - 2\eta \frac{M\tilde{\Gamma}_\text{meas}^2 \sin^2\theta}{\left(M\tilde{\Gamma}_\text{meas} + \frac{4\Omega_\text{m}\bar{n}_m}{Q_m} \right)^2} \left(1+ \frac{4\Omega_\text{m}\bar{n}_m}{M\tilde{\Gamma}_\text{meas}Q_m}\right)$, at the optimal frequency $\delta_\text{opt}\sim \left(M\tilde{\Gamma}_\text{meas} + \frac{4\Omega_\text{m}\bar{n}_m}{Q_m}\right) \frac{\cos\theta}{\sin\theta}$. In the strong coupling regime, $M\tilde{\Gamma}_\text{meas} \gg \frac{4\Omega_\text{m}\bar{n}_m}{Q_m}$, the minimal variance, $V^{(M)}_\text{min} \approx 2 -\frac{2\eta}{M}\sin^2\theta$, is independent of the optomechanical coupling strength, $G$. On the contrary, in the weak coupling regime, $M\tilde{\Gamma}_\text{meas} \ll \frac{4\Omega_\text{m}\bar{n}_m}{Q_m}$, the minimal variance, $V^{(M)}_\text{min} \approx 2 -\frac{\eta Q_m M \tilde{\Gamma}_\text{meas}}{2\Omega_\text{m}\bar{n}_m}\sin^2\theta$, is proportional to the square of optomechanical coupling strength, $G^2$.

At $\theta=\pm \pi/4$ the variance becomes
\begin{equation} \label{eq:SmpV}
  V^{(M)}_{jl}(\omega) \approx \left[2 + \eta M\left(\frac{\tilde{\Gamma}_\text{meas}}{\delta}\right)^2  + \eta \frac{\tilde{\Gamma}_\text{meas}}{\delta}(4\frac{\Omega_m}{\delta}\frac{\bar{n}_m}{Q_m} \mp 2) \right] (2\bar{n}_\text{o} +1)\;.
\end{equation}
The bipartite entanglement of CV states requires $|\delta| > M\tilde{\Gamma}_\text{meas}/2 + \Omega_\text{m}\bar{n}_m/Q_m$. At the optimal frequency $|\delta_\text{opt}|= M\tilde{\Gamma}_\text{meas} + 4 \Omega_\text{m0}\frac{\bar{n}_\text{th}}{Q_\text{m}}$, we obtain the minimal variance, $V_\text{min} \approx \left( 2- \frac{\eta\tilde{\Gamma}_\text{meas}}{M \tilde{\Gamma}_\text{meas} + 4 \Omega_{\text{m}0} \frac{\bar{n}_m}{Q_m}}\right)(2\bar{n}_\text{o} +1)>\left(2-\frac{\eta\tilde{\Gamma}_\text{meas}}{M}\right)$.

\FloatBarrier

%


\end{document}